\documentclass[aps,prd,preprint,groupedaddress,showpacs]{revtex4}
\usepackage{amssymb,amsmath,graphicx}

\usepackage{verbatim,color,ulem}

\begin{document}
\title{Recoherence by Squeezed States in Electron Interferometry}
\author{Jen-Tsung Hsiang}
\email{cosmology@mail.ndhu.edu.tw}
\affiliation{Department of Physics, National Dong Hwa University,
Hualien, Taiwan, R.O.C.}
\author{L. H. Ford}
\email{ford@cosmos.phy.tufts.edu}
\affiliation{Institute of Cosmology, Department of Physics, Tufts
University, Medford, Massachusetts 02155, USA.}

\begin{abstract}
Coherent electrons coupled to the quantized electromagnetic field undergo decoherence which
can be viewed as due either to fluctuations of the Aharonov-Bohm phase or to photon emission.
When the electromagnetic field is in a squeezed vacuum state, it is possible for this
decoherence to be reduced, leading to the phenomenon of recoherence. This recoherence 
effect requires 
electrons which are emitted at selected times during the cycle of the excited mode
of the electromagnetic field. We show that there are bounds on the degree of recoherence
which are analogous to quantum inequality restriction on negative energy densities in quantum
field theory. We make some estimates of the degree of recoherence, and show that although
small, it may be observable.
\end{abstract}

\pacs{03.65.Yz, 03.75.-b, 41.75.Fr, 42.50.Lc}
\maketitle

\baselineskip=13pt

\section{Introduction}

\begin{figure}
\begin{center}
    \scalebox{0.5}{
        \includegraphics{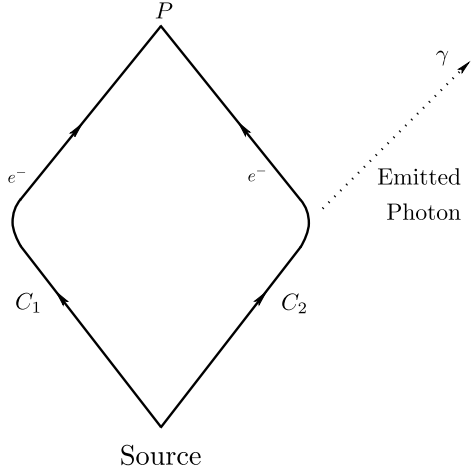}}
        \caption{An electron interference experiment in which the electrons may take either
one of two paths, $C_1$ or $C_2$, from the source to the point $P$ where the interference pattern
is formed. The emission of photons by the electrons tends to cause decoherence. The detection
of an emitted photon with wavelength smaller than the path separation can reveal which path
a particular electron takes, and hence causes decoherence. }
\label{fig:paths}
\end{center}
\end{figure}

The interference of electrons is one of the most basic phenomena which illustrate the
quantum nature of electrons. In recent years, technological advances have allowed electron 
interferometry to be used for a variety of investigations~\cite{aT1982,TEMKE1989,NH93,T05,SH07}. 
However, the quality of the
interference pattern obtained with electrons is never as good as can be achieved with light
or with neutral atoms. This can be attributed to the fact that charged particles interact
more strongly with their environment  than do
photons or neutral atoms, and are hence more
subject to loss of quantum coherence, or decoherence. This can arise from a variety of effects,
such as interaction with random fields in the interferometer or with
thermal radiation. Recently, Sonentag and Hasselbach~\cite{SH07} observed
decoherence as a result of dissipative interaction with image charge
fields near an imperfectly conducting plate.  In
principle, these effects could be removed if there are no photons or classical fields in
the interferometer. However, there is still a decoherence effect even when the quantized 
electromagnetic field is initially in its vacuum 
state~\cite{SAI1990,SAI1991,lF1993,lF1994,lF1997,BP2001,MPV2003}. 
This effect can be interpreted as arising
from photon emission by the electrons. The emission of a photon with sufficiently short
wavelength can reveal which path a particular electron takes and hence acts to destroy the
interference pattern. (See Fig.~\ref{fig:paths}.) An equivalent description is in terms of
a fluctuating Aharonov-Bohm phase.

In this paper, we will be concerned with the effects of squeezed photon states on the 
electron coherence. Squeezed states describe reduced quantum fluctuations in one variable
at the expense of increased fluctuations in the conjugate variable. One remarkable property
is that they can exhibit locally negative energy densities. This phenomenon can be understood
as a suppression of vacuum fluctuations. The normal ordered stress tensor operator is a
difference between an expectation value in a given state and that in the vacuum, a difference
which can become negative. As will be detailed in the next section, electron decoherence
due to a fluctuating electromagnetic field can be ascribed to fluctuations of the Aharonov-Bohm 
phase. As we will demonstrate, it is possible to use squeezed states of the quantized
electromagnetic field to reduce these fluctuations, leading to a decrease in decoherence, which
we will call ``recoherence''. Squeezed states and coherent electrons
were discussed in a somewhat different context by Vourdas and Sanders~\cite{VS98}, who 
developed a procedure by which coherent electrons may be used to
measure quantum states of the electromagnetic field.

The outline of this paper is as follows: In Sect.~\ref{S:general_formalism}, we outline
the formalism of decoherence by a fluctuating Aharonov-Bohm phase. We then apply this
formalism to the case of a single-mode squeezed state in Sect.~\ref{S:single_mode} and 
calculate the degree of recoherence which is possible. These calculations are extended to
multi-mode squeezed states in Sect.~\ref{S:multiple_mode}, and some
numerical estimates are given in Sect.~\ref{sec:estimates}. In Sect.~\ref{sec:final}, we
summarize and discuss our results. Some of the properties of squeezed states are reviewed
in the Appendix. Unless otherwise noted, we use Lorentz-Heaviside units with $\hbar = c =1$. 

\section{general formalism}\label{S:general_formalism}

Here we briefly review the effects of electromagnetic field fluctuations on electron 
coherence~\cite{SAI1990,SAI1991,lF1993,lF1994,lF1997,BP2001,MPV2003,VS98,jH2004,jH2004a,HL06,HL08}.
 Consider a double slit interference experiment in which coherent electrons
can take either one of two paths, as illustrated in Fig.~\ref{fig:paths}. First consider the 
case of no field fluctuations.  If the amplitudes
for the electrons to take path $C_1$ and $C_2$ are $\psi_1$ and $\psi_2$, respectively,
 to point $P$, then the mean number of electrons at $P$ will be proportional to
\begin{equation}
n(P)= |\psi_1+\psi_2|^2 = |\psi_1|^2+|\psi_2|^2 + 2 {\rm Re}(\psi_1\, \psi_2^*) \,.
\end{equation}  
In the presence of a classical, non-fluctuating electromagnetic field described by vector 
potential $A^\mu$, there will be an Aharonov-Bohm phase shift of the form~\cite{AB1959}. 
\begin{equation}
\varphi_{AB} = e \oint_C  d x^\mu \, A_\mu \,,
\end{equation}
where the integral is taken around the closed path $C= C_1 -C_2$. 
This shifts the locations
of the interference minima and maxima, but does not alter their relative amplitudes, the
contrast. 

If the electromagnetic field undergoes fluctuations, then the situation is different. In 
this case, the fluctuating  Aharonov-Bohm phase causes a change in the contrast by 
a factor of
\begin{equation}
\Gamma =  {\rm e}^W \,,
\end{equation}
where we define the {\it coherence functional} by
\begin{equation}
W =-\frac{1}{2} \langle \varphi_{AB}^2 \rangle 
\end{equation}
with the angular brackets denoting averaging over the fluctuations. This functional
can be expressed as
\begin{equation}\label{E:gamma}
    W=-2\pi\alpha\oint_Cdx_{\mu}\oint_Cdx'_{\nu}\;D^{\mu\nu}(x,x')\,,
\end{equation}
where $\alpha$ is the fine-structure constant and
\begin{equation}\label{E:hada}
    D^{\mu\nu}(x,x')=\frac{1}{2}\,\bigl\langle \bigl\{A^{\mu}(x),A^{\nu}(x')
          \bigr\} \bigr\rangle\,.
\end{equation}
So far, we have not specified the source of the fluctuations, which could be thermal,
quantum, or due to averaging over classical time variations~\cite{jH2004a}. In this paper, we will be
concerned with quantum fluctuations in a squeezed vacuum state.

\section{single-mode squeezed vacuum}\label{S:single_mode}
\subsection{Renormalized Coherence Functional, $W_R$}

In this section, we consider the special case where the quantized electromagnetic field
is in a state in which one mode is excited to a squeezed vacuum state, and all other
modes remain in the ground state. We take the excited mode to be a
plane wave in a box with periodic boundary conditions,
 with wave vector $\bar{\mathbf{k}}$ and polarization $\bar{\lambda}$,
 so the quantum state may be denoted by $|\zeta_{\bar{\lambda}\bar{\mathbf{k}}} \rangle$. 
The paths  $C_{1}$ and $C_{2}$ are taken to be in the $xz$ plane and we
suppose that the electron wavepacket is prepared to be highly localized about the 
classical trajectory and its dispersion can be ignored in the classical limit~\cite{lF1993}.
 If the $x$-component of the electron velocity is constant, and the trajectories $C_{1}$ 
and $C_{2}$ are chosen to be
symmetric to one another with respect to the $z=0$ plane, then in a
co-moving frame where the electron only has the sideways motion along
the $z$ axis, the quantity $W$ can be greatly simplified to
\begin{equation}
    W=-2\pi\alpha\oint_Cdz\oint_Cdz'\;D^{zz}(x,x')\,.
\end{equation}

If we are only interested in the change of the fringe contrast due to the excitation of a 
particular squeezed vacuum mode, then after subtracting the vacuum contribution of 
all modes, we have the renormalized coherence functional given by
\begin{equation}\label{E:W_R}
    W_{R}=-\pi\alpha\oint_Cdz\oint_Cdz'\;\bigl<\zeta_{\bar{\lambda}\bar{\mathbf{k}}}
\big|\bigl\{A^{z}\left(x\right),A^{z}\left(x'\right)\bigr\}
\big|\zeta_{\bar{\lambda}\bar{\mathbf{k}}}\bigr>_{R}\,.
\end{equation}
Here we use the subscript $R$ to denote the renormalized quantity, which has the Minkowski
vacuum term subtracted. In the Coulomb
 gauge, the $z$ component of the vector potential in
the plane wave expansion takes the form
\begin{equation}
    A^{z}\left(x\right)=\frac{1}{\sqrt{V}}\sum_{\mathbf{k}}\frac{1}{\sqrt{2\omega}}
\sum_{\lambda=1}^2\mathbf{e}_z\cdot\boldsymbol{\varepsilon}_{\lambda}
\left(\mathbf{k}\right)\left(a_{\lambda\mathbf{k}}^{\vphantom{\dagger}}
e^{-ik\cdot x}+a_{\lambda\mathbf{k}}^{\dagger}e^{ik\cdot x}\right)
\end{equation}
where $\mathbf{e}_z$ is the unit vector along the $z$ axis and 
$\boldsymbol{\varepsilon}_{\lambda}$ are unit polarization vectors. The quantity 
$V$ is the box normalization volume and $\omega=\left|\mathbf{k}\right|$. If we 
further assume that the mode $(\bar{\lambda},\bar{\mathbf{k}})$ is polarized in the 
$z$ direction, and its wave vector is directed in the $y$ direction, then the renormalized
 Hadamard function in squeezed vacuum is given by
\begin{equation}
    \bigl<\zeta_{\bar{\lambda}\bar{\mathbf{k}}}\big|\bigl\{A^{z}\left(x\right),A^{z}
\left(x'\right)\bigr\}\big|\zeta_{\bar{\lambda}\bar{\mathbf{k}}}\bigr>_R=
\frac{1}{V\bar{\omega}}\left[-\mu\nu\,e^{-i\bar{\omega}\left(t+t'\right)}+
\left|\nu\right|^2e^{-i\bar{\omega}\left(t-t'\right)}+\text{C.C.}\right]\,,
\end{equation}
where $\mu=\cosh r$, $\nu=e^{i\,\theta}\sinh r$, and 
$\zeta_{\bar{\lambda}\bar{\mathbf{k}}}=r\,e^{i\theta}$ is the complex squeeze parameter defined in
the Appendix. Thus Eq.~\eqref{E:W_R} becomes
\begin{equation}
    W_R=-\frac{\pi\alpha}{V\bar{\omega}}\oint_Cdz\!\oint_Cdz'\,
\left[-\mu\nu\,e^{-i\bar{\omega}\left(t+t'\right)}+\left|\nu\right|^2
e^{-i\bar{\omega}\left(t-t'\right)}+\text{C.C.}\right]\,.
\end{equation}
Thus, this is the factor that accounts for the contrast change of the electron 
interference fringe in the present arrangement, where we shine a polarized beam in a single mode
squeezed vacuum state in the direction perpendicular to the plane of the electron paths.

To evaluate $W_R$, we pick a path which is twice differentiable,
\begin{equation}
    z\left(t\right)=\frac{R}{T^4}\left(t^2-T^2\right)^2 \, ,
\end{equation}
where $2T$ and $2R$ can be thought of as the effective flight time and path separation,
 respectively. Electrons which start from the source at different times will experience
different fluctuations. We can study this effect by letting $t_0$ be the electron emission
time, in which case the quantity $W_R$ becomes
\begin{align}\label{E:wr}
    W_R&=-\frac{4\pi\alpha}{V\bar{\omega}}\int_{-T}^{T}dt\!
\int_{-T}^{T}dt'\;v^{\vphantom{'}}_zv'_z\left[-\mu\eta\,
e^{-i\,\bar{\omega}\left(t+t'\right)-i\,2\bar{\omega}t_0+i\,\theta}+\eta^2
e^{-i\,\bar{\omega}\left(t-t'\right)}+\text{C.C.}\right]\notag\\
&=-\frac{8\pi\alpha\, \eta}{V\bar{\omega}}
\Bigl[\mu\,\cos(2\bar{\omega}t_0-\theta)+\eta \Bigr]M\,,
\end{align}
with $v_z=dz(t)/dt$, $\eta=\sinh r$, and
\begin{equation}
	 M=\left(\frac{16R}{\bar{\omega}^4T^4}\right)^2
\Bigl[(-3+\bar{\omega}^2T^2)\sin\bar{\omega}T+3\bar{\omega}T\cos\bar{\omega}T
\Bigr]^2\,.
\end{equation}
The quantity $M$ does not depend on the electron emission time $t_0$ and is always 
positive definite, so the sign of $W_R$ is solely determined by the quantity
$\mu\,\cos(2\bar{\omega}t_0-\theta)+\eta$.

\subsection{Interpretation of $W_R$}
An intriguing feature is that the values of $W_{R}$ are not always negative, and they 
can be positive, depending on the parameters $t_0$, $\mu$ and $\eta$. It implies that 
the amplitude factor $e^{W_R}$ may be larger than unity for some moments, which in 
turn means that the contrast on the screen can be higher than it would otherwise be
 for the vacuum state. This is generally interpreted as enhancement of
 coherence, or recoherence. This
 contrast change can not be observed right away when only one electron is released 
at each moment. We have to wait for sufficiently long time so that enough electrons 
are accumulated to have visible patterns. However, since $t_0$ is related to the electron
 emission time and is assumed to be a random variable, if the time scale of the 
measurement is much longer than the flight time $2T$, then it is an long-time-averaged
 result that should be observed,
\begin{equation}
    \overline{W}_R\equiv\lim_{\Xi\to\infty}\frac{1}{2\Xi}\int_{-\Xi}^{\Xi}dt_0\;W_R=
-\frac{8\pi\alpha}{V\bar{\omega}}\,\eta^{2}M<0\,.
\end{equation}
Hence this time-averaged value of $W_R$ is always negative. This means that measurements
which average over a long time will always find decoherence from the presence of the squeezed
vacuum state, but measurements on shorter time scales have a chance to find $W_R >0$,
which means transient recoherence. 

This feature bears a strong resemblance to the issue of negative energy density. It 
is known that quantum field theory has the remarkable property that local energy 
density can be negative even though the energy density is a positive-definite quantity 
in classical physics. It is a general feature of both free and interacting theories that 
there exist states in which the energy density at a particular point can be arbitrarily 
negative~\cite{EGJ1965}. Nonetheless, the total energy, integrated over all space, is 
required to be non-negative. It is also shown that there exist quantum 
inequalities~\cite{lF1978,Ford91,FR97,FE98} 
which constrain the magnitude and duration of the negative
 energy density and flux. Physically, the inequalities imply that the energy density seen 
by an observer cannot be arbitrarily negative for an arbitrarily long
period of time. Marecki~\cite{Marecki02,Marecki08} has recently derived
variants of the quantum inequalities for limiting the amount of
squeezing which might be observed in photodetection experiments in
quantum optics.
Therefore it is interesting to know whether there exists a similar inequality on the 
quantity $W_R$, at least for the squeezed vacuum, to limit how positive it can be and 
for how long.

\begin{figure}
\begin{center}
    \scalebox{1.2}{
        \includegraphics{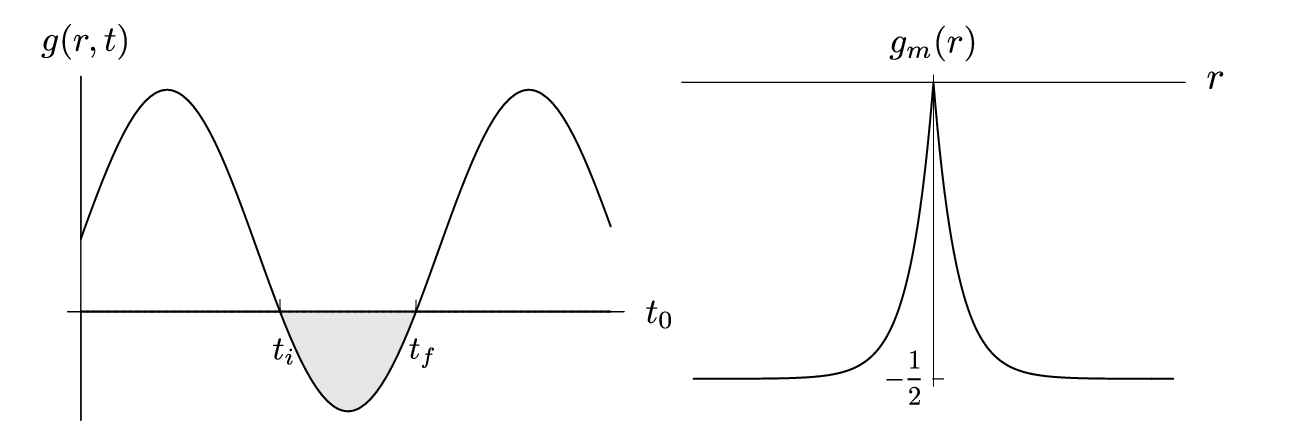}}
        \caption{The left figure shows the behavior of $g\left(r,t\right)$ defined in
     Eq.~(\ref{eq:g}), as a function of the emissions time
 $t_0$. The right figure shows how how the minimum value of $g$ as a function of $t_0$, 
$g_m\left(r\right)$, depends on $r$.}
\label{Fi:squeezed_state}
\end{center}
\end{figure}

Define a new function $g\left(r,t\right)$ which includes all $r$-dependence of
 the quantity $W_R$,
\begin{equation}
    g\left(r,t\right)=\eta\left[\mu\cos\left(\alpha+\beta t_0\right)+\eta\right],
   \label{eq:g}
\end{equation}
with
\begin{equation}
    \alpha=\bar{\omega}T-\theta\,,\qquad\beta=2\bar{\omega}\,.
\end{equation}
Then $W_R$ can be expressed in terms of $g\left(r,t\right)$ by
\begin{equation}
    W_R=-\frac{8\pi\alpha}{V\bar{\omega}}M\,g\left(r\right)\,,
\end{equation}
and the behavior of $g\left(r,t\right)$ will tell us how the quantity $W_R$ depends on
 the parameter $r$. From Fig.~\ref{Fi:squeezed_state}, we see that $W_R$ is positive 
only if the time variable $t_0 \bmod \pi/\omega$ lies in the time interval between 
$t_i$ and $t_f$. If we can somehow collect only those electrons that are emitted during 
those moments, then we can guarantee a positive average value of the quantity $W_R$.
 Hence the recoherence of the electron interference may be maintained and may remain
 strong enough to be observed. Next, we will discuss how to compute the averaged 
value, $\widetilde{W}_R$, formed by averaging $W_R$ over the interval in which it is
positive.

\subsection{Behavior of $g\left(r,t\right)$}\label{S:g_fn}

Since only the function $g\left(r,t\right)$ will affect the overall sign of $W_R$, it is 
sufficient to calculate the time averaged value of $g\left(r,t\right)$ between $t_i$ and $t_f$. 
Then $\widetilde{W}_R$ is just proportional to this averaged value $\widetilde{g}(r)$. 
Here we note that $g\left(r,t\right)$ is only defined for $r \geq 0$. From
 Fig.~\ref{Fi:squeezed_state}, the condition $g\left(r,t\right)=0$ is satisfied when $t_0$
 is equal to either $t_i$ or $t_f$, and $g\left(r,t\right)$ is symmetric about those 
values of $t_0$ that satisfy $\alpha+\beta t_0=(2n+1)\pi$, with $n$ an integer. Thus let
\begin{equation}\label{E:t2}
    \pi-\xi=\alpha+\beta t_i\,,\qquad\pi+\xi=\alpha+\beta t_f\,,
\end{equation}
where $t_i$ is assumed to be smaller than $t_f$. Then it is easy to see that $\xi$ 
satisfies
\begin{equation}
    \mu\cos\left(\pi\pm\xi\right)+\eta=0\qquad\Rightarrow\qquad
\cos\xi=\frac{\eta}{\mu}<1\,,
\end{equation}
and from Eq.~(\ref{E:t2}), we have
\begin{equation}
    \Delta t=t_f-t_i=\frac{2\xi}{\beta}=\frac{2}{\beta}
     \cos^{-1}\left(\frac{\eta}{\mu}\right) \,.
\label{E:dt_r_pos}
\end{equation}
On the other hand, the integration of $\cos\left(\alpha+\beta t_0\right)$ over $t_0$ 
between $t_i$ and $t_f$ yields
\begin{equation}
    \int_{t_i}^{t_f}dt_0\,\cos\left(\alpha+\beta t_0\right)=-\frac{2}{\beta\mu}\,.
\end{equation}
Putting the above results together, we have that the time average of the quantity 
$g\left(r,t\right)$ over the interval between $t_i$ and $t_f$ is given by
\begin{align}
    \widetilde{g}\left(r\right)&=\frac{1}{t_f-t_i}\int_{t_i}^{t_f}dt_0\,\eta
\left[\mu\cos\left(\alpha+\beta t_0\right)+\eta\right]\\
                            &=-\frac{\eta}{\cos^{-1}({\eta}/{\mu})}+\eta^2\,,
\label{E:g_r_pos}
\end{align}
and thus the average of $W_R$ over the same interval is
\begin{equation}\label{E:ave_W_R}
    \widetilde{W}_R=-\frac{8\pi\alpha}{V\bar{\omega}}M\eta
\left[-\frac{1}{\cos^{-1}({\eta}/{\mu})}+\eta\right]\,.
\end{equation}
In addition, the knowledge of the local extrema of the function $g\left(r,t\right)$ with 
respect to $t_{0}$ will prove useful. Its local minimum along the $t_0$ axis is 
given by
\begin{equation}\label{E:gm_r_pos}
    g_m\left(r\right)=\eta\left(-\mu+\eta\right)=-\frac{1}{2}\left(1-e^{-2r}\right)\,,
\end{equation}
while the local maximum value of $g\left(r,t\right)$ is
\begin{equation}\label{E:gM_r_pos}
    g_M\left(r\right)=\eta\left(\mu+\eta\right)=\frac{1}{2}\left(e^{2r}-1\right)\,.
\end{equation}
\begin{figure}
\begin{center}
    \scalebox{1.2}{
        \includegraphics{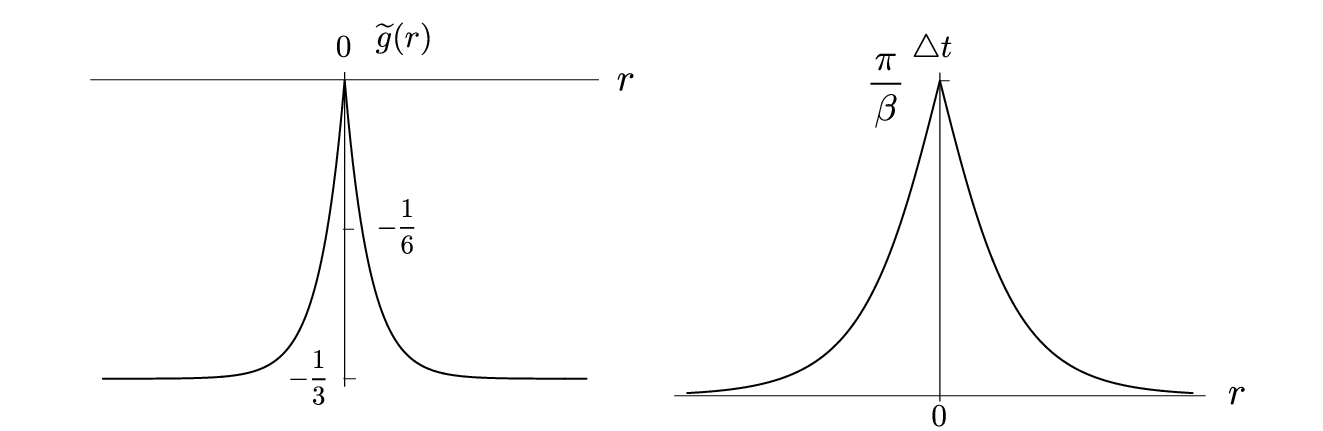}}
        \caption{The left figure shows how $\widetilde{g}\left(r\right)$, the average of
$g(r,t)$ over the interval when $g(r,t) < 0$, as a function of $r$. The right figure
illustrates the width of this interval, $\triangle t$, also as a function of $r$.}
\label{Fi:squeezed_state1}
\end{center}
\end{figure}
Because the function $\widetilde{g}\left(r,t\right)$ and the extrema of $g\left(r\right)$ 
are monotonic functions of $r$, we may consider only two limiting values of $r$. In the 
limit that the parameter $r$ approaches positive infinity, we have
\begin{align}
    r\rightarrow+\infty\qquad\qquad\qquad g_m\left(r\right)&\approx
-\frac{1}{2}+\mathcal{O}\left(e^{-2r}\right)\,,&g_M\left(r\right)&\approx
\frac{1}{2}\,e^{2r}+\mathcal{O}\left(1\right)\,,\\
    \Delta t&\approx\frac{4}{\beta}\,e^{-r}+\mathcal{O}\left(e^{-3r}\right)\,,&
\widetilde{g}\left(r\right)&\approx-\frac{1}{3}+\mathcal{O}\left(e^{-2r}\right)\,.
\end{align}
On the other hand, when the parameter approaches to $0^+$, we have
\begin{align}
    r\rightarrow0^+\qquad\qquad\qquad   g_m\left(r\right)&\approx
-r+\mathcal{O}\left(r^2\right)\,,&g_M\left(r\right)&\approx
 r+\mathcal{O}\left(r^2\right)\,,\\
    \Delta t&\approx\frac{\pi}{\beta}+\mathcal{O}\left(r\right)\,,&
\widetilde{g}\left(r\right)&\approx-\frac{2}{\pi}r+\mathcal{O}\left(r^2\right)\,.
\end{align}
Thus when $r$ gradually goes to zero, both the maximum and the minimum of the 
function $g\left(r,t\right)$ goes to zero from above and below respectively. We 
intermediately know that $g\left(r,t\right)$ will be identically equal to zero in this limit. 
The width of the interval, $\Delta t$, approaches to a finite value $\pi/\beta$. Thus the
 average value $\widetilde{g}\left(r\right)$ will be vanishing accordingly.

In contrast, the maximum of the function $g\left(r,t\right)$ grows 
exponentially as $r$ increases, and the minimum decreases, approaching a
lower bound of $-1/2$. 
The width of the interval, over which the average is performed, decreases to zero in 
the limit $r\rightarrow\infty$. Nonetheless, the average value 
$\widetilde{g}\left(r\right)$ remains finite and is equal to $-1/3$ in this limit. This is
 the lower bound of the function $\widetilde{g}\left(r\right)$.

\subsection{Bound on recoherence and preservation of unitarity}

In short, the function $g\left(r,t\right)$ is always bounded from below by a finite value 
of $-1/2$ while it is unbounded above. 
 Furthermore, although the width of the integration interval vanishes as 
$r \rightarrow \infty$, the function $\widetilde{g}\left(r\right)$ is 
still bounded between $0$ and $-1/3$. Thus we can see that there does exist a upper 
bound for the average value $\widetilde{W}_R$  given by
\begin{align}
    \max\left[\widetilde{W}_R\right]&=\frac{8\pi\alpha}{3V\bar{\omega}}\,M\notag\\
    &=\frac{8\pi\alpha}{3V\bar{\omega}}
\left(\frac{16R}{\bar{\omega}^4T^4}\right)^2
\Bigl[(-3+\bar{\omega}^2T^2)\sin\bar{\omega}T+
3\bar{\omega}T\cos\bar{\omega}T\Bigr]^2\,.
                   \label{eq:maxW}
\end{align}

We see that more often than not, $W_R < 0$, meaning that the photons in the squeezed
state tend to increase decoherence above what is already present. However, by forming
an interference pattern with carefully selected electron emitted in certain time intervals,
we can reverse this tendency, and attain positive values of $W_R$, leading to
recoherence. Note that the recoherence effect is maximal when $r$ is large, 
corresponding to a large mean number of photons in the squeezed vacuum state , given 
by $\bar{n}=\eta^{2}=\sinh^{2}r$.

One might be concerned that $W_R > 0$ could lead to a violation of unitarity, but this
is not the case, because the vacuum effect will always dominate and lead to
$W = W_0 +W_R <0$.  It suffices to compute this combined contribution of one 
mode $(\bar{\lambda},\bar{\mathbf{k}})$ to $W$. It is straightforward to find 
that, for this mode, $W_{0}(\bar{\lambda},\bar{\mathbf{k}})$ is given by
\begin{equation}
	W_{0}(\bar{\lambda},\bar{\mathbf{k}})=-\frac{4\pi\alpha}{V\bar{\omega}}\,M
\end{equation}
for the same path configuration; while the  maximal value of $W_{R}$ is
\begin{equation}
	W_{R}(\bar{\lambda},\bar{\mathbf{k}})=\frac{8\pi\alpha}{3V\bar{\omega}}\,M\,.
\end{equation}
Therefore, we have the combined value of $W$ for this mode
  $(\bar{\lambda},\bar{\mathbf{k}})$ is negative:
\begin{equation}\label{E:tot}
	 W(\bar{\lambda},\bar{\mathbf{k}})=
W_{0}(\bar{\lambda},\bar{\mathbf{k}})+W_{R}(\bar{\lambda},\bar{\mathbf{k}})
=-\frac{4\pi\alpha}{3V\bar{\omega}}\,M<0\,.
\end{equation}
Since for the rest of the modes 
$(\lambda,\mathbf{k})\neq(\bar{\lambda},\bar{\mathbf{k}})$,  we have 
$W(\lambda,\mathbf{k})=W_{0}(\lambda,\mathbf{k})<0$, which in turn implies that
\begin{equation}
	\sum_{\mathbf{k},\,\lambda}W(\lambda,\mathbf{k})<0\,.
\end{equation}

\section{multi-mode squeezed vacuum of finite bandwidth}
\label{S:multiple_mode}

So far, we have considered a single excited mode. Now we wish to extend our result
to the case of many excited modes.    Assume that the electromagnetic field is initially 
prepared in the state
\begin{equation}
    |\nu\rangle= |0_1\rangle \cdots |0_{r}\rangle |\zeta_{r+1}\rangle
\cdots |\zeta_{r+n}\rangle |0_{r+n+1}\rangle \cdots,
\end{equation}
where $|\zeta_i\rangle$ is the squeezed vacuum state for mode $i$. 
Thus $n$ modes, $r+1$ to $r+n$, are excited 
in squeezed vacuum states and the rest remain in the vacuum state. Here the 
subscripts in the bra and ket denote the mode labels. 
We assume that the excited modes are all linearly polarized in the
same direction. The distribution of the wave vectors for the
excited modes are also assumed to be sharply centered about some wave vector
 $\bar{\mathbf{k}}$, which
is parallel to the $y$-axis, so that the distribution forms a small
cone with a solid angle $d\Omega$ about $\bar{\mathbf{k}}$, and
coherence is maintained among these modes. Thus the quantity $W_R$ is then given by
\begin{align}
    W_R &=-\frac{2}{V}\,e^2\sum_{\mathbf{k}}\omega
\left[\mu_{\omega}\eta_{\omega}\cos\left(2\omega t_0-\theta_{\omega}\right)+
\eta_{\omega}^2\right]\\
        &\qquad\qquad\qquad\times\left(\frac{16R}{\omega^3T^2}\right)^2
\left[\sin\omega T+\frac{3\cos\omega T}{\omega T}-
\frac{3\sin\omega T}{\omega^2T^2}\right]^2,
\end{align}
where $\mathbf{k}\in\{\mathbf{k}_{r+1}\ldots \mathbf{k}_{r+n}\}$ and
 $\omega=|\mathbf{k}|$. If the distribution of modes is dense enough, then it can be
 described by a smooth mode-distribution function $f\left(\mathbf{k}\right)$, centered
 at $\bar{\mathbf{k}}$. If we further assume that $f\left(\mathbf{k}\right)$ depends 
only on the frequency $\omega$, we can rewrite the mode summation as an integration 
over the phase space volume,
\begin{equation}
    \frac{1}{V}\sum_{k}=\int\frac{d^3k}{(2\pi)^3}=
\frac{d\Omega}{(2\pi)^3}\int_0^{\infty}d\omega\,\omega^2f\left(\omega\right)
\end{equation}
where $f\left(\omega\right)$ is the mode-distribution function,
peaked at $\omega = \bar{\omega}$ with the width $\Delta\omega$. If
$\Delta \omega \ll \bar{\omega}$, so the bandwidth is not overly 
wide, we may assume the squeeze parameters $r_{\omega}$, $\theta_{\omega}$, 
$\mu_{\omega}$ and $\eta_{\omega}$ are constants, independent of frequency for all 
excited modes within the band, thus removing the subscript $\omega$ from now on. 
Therefore, $W_R$ becomes
\begin{align}
    W_R &=-2\,e^2\left(\frac{16R}{T^2}\right)^2\frac{d\Omega}{(2\pi)^3}
\int_0^{\infty}d\omega\;f\left(\omega\right)
\left[\mu\eta\cos\left(2\omega t_0-\theta\right)+\eta^2\right]\\
        &\qquad\qquad\qquad\times\,\frac{1}{\omega^3}\left[\sin\omega T+
\frac{3\cos\omega T}{\omega T}-\frac{3\sin\omega T}{\omega^2T^2}\right]^2\,.
\end{align}

If we only sample electrons which will contribute to recoherence, then according to 
the discussion in the previous section, the expression 
$\mu\eta\, \cos (2\omega t_0-\theta)+\eta^2$ is  replaced by 
$\widetilde{g}(r)$, given by Eq.~(\ref{E:g_r_pos}), which is
independent of frequency. Note that time interval $\Delta t$ over
which we sample is inversely proportional to $\bar{\omega}$, but the
bandwidth $\Delta \omega$ is independent of  $\Delta t$, so long as 
$\Delta \omega \ll \bar{\omega}$. 
Moreover, if we assume that the 
mode-distribution function takes the form
\begin{equation}
    f\left(\omega\right)=\begin{cases}
                                    1,\qquad\text{if  }\bar{\omega}-
\Delta\omega\leq\omega\leq\bar{\omega}+\Delta\omega\,,\\
                                    0,\qquad\text{otherwise}\,,
                                \end{cases}
\end{equation}
then the quantity $\widetilde{W}_R$ reduces to
\begin{equation}
    \widetilde{W}_R=-2\,e^2\!\left(\frac{16R}{T^2}\right)^2\!
\overline{g}\left(r\right)\!\frac{d\Omega}{(2\pi)^3}\!
\int_{\bar{\omega}-\Delta\omega}^{\bar{\omega}+\Delta\omega}\!\!d\omega\,
\frac{1}{\omega^3}\left[\sin\omega T+\frac{3\cos\omega T}{\omega T}-
\frac{3\sin\omega T}{\omega^2T^2}\right]^2\,.
\end{equation}
In principle, the integration on the right-hand side can be carried out exactly; however,
 for simplicity, we only show the result of the integral to the order
$\mathcal{O}\left(\Delta\omega/\bar{\omega}\right)$,
\begin{align}
    &\int_{\bar{\omega}-\Delta\omega}^{\bar{\omega}+\Delta\omega}\!\!d\omega\,
\frac{1}{\omega^3}\left[\sin\omega T+\frac{3\cos\omega T}{\omega T}-
\frac{3\sin\omega T}{\omega^2T^2}\right]^2\\
    =&\frac{2}{\bar{\omega}^6T^4}\left[\bar{\omega}^2T^2\sin\bar{\omega}T+
3\bar{\omega}T\cos\bar{\omega}T-3\sin\bar{\omega}T\right]^2
\frac{\Delta\omega}{\bar{\omega}}+\mathcal{O}\left(\frac{\Delta\omega^2}
{\bar{\omega}^2}\right)\,,
\end{align}
where we have assumed $\Delta\omega\, T\ll1$ and
$\Delta\omega/\bar{\omega} \ll1$. 
Therefore, the leading contribution of $\widetilde{W}_R$ is given by
\begin{equation}\label{E:temp7}
    \widetilde{W}_R=-\,e^2\frac{R^2}{T^2}\,
\overline{g}\left(r\right)\frac{d\Omega}{(2\pi)^3}
\left(\frac{32}{\bar{\omega}^3T^3}\right)^2
\left[\bar{\omega}^2T^2\sin\bar{\omega}T+
3\bar{\omega}T\cos\bar{\omega}T-3\sin\bar{\omega}T\right]^2
\frac{\Delta\omega}{\bar{\omega}}\,.
\end{equation}

\section{Some Numerical Estimates}
\label{sec:estimates}

\subsection{Single Mode in a Cavity}

Our treatment of a single excited mode in Sect.~\ref{S:single_mode} assumed periodic boundary
conditions for simplicity. However, the result should be useful for making an order-of-magnitude 
estimate of the effect in a cavity with more realistic boundary conditions. First define the
function
\begin{equation}
F(x) = \left(\frac{32}{x^3}\right)^2\, [(x^2-3) \sin x +3x \cos x ]^2 \,.  \label{eq:F}
\end{equation}
This function has a maximum value of $F(3.34) \approx 96.4$ at $x \approx 3.34$, and for
large arguments is approximately
\begin{equation}
F(x) \approx \frac{1024}{x^2} \, \sin^2 x \,, \qquad x \gg 1 \,. \label{eq:Flarge}
\end{equation}
Let $\lambda = 2\pi/\bar{\omega}$ be the wavelength of the excited mode. If we assume that
the averaged coherence functional, $\widetilde{W}_R$, attains its maximum value given in 
Eq.~(\ref{eq:maxW}), then we can express this value as
\begin{equation}
\widetilde{W}_R \approx \frac{\alpha}{12 \pi^2 }\, \frac{\lambda^3}{V} \, 
\left(\frac{R}{T}\right)^2\, F(2\pi T/\lambda) \,.
\end{equation}
If we assume $2\pi T \gg \lambda$, and  use the large argument form for $F$, Eq.~(\ref{eq:Flarge}), 
we can write
\begin{equation}
\widetilde{W}_R \approx 8 \times 10^{-4}\, \frac{\lambda^3}{V} \, \left(\frac{R}{T}\right)^2\,
 \left(\frac{\lambda}{T}\right)^2\,.
\end{equation}
For a rough estimate, let us take $V \approx \lambda^3$ and $R \approx \lambda$, corresponding to
the lowest frequency mode in the cavity and a path separation of the order of the cavity size.
This leads to
\begin{equation}
\widetilde{W}_R \approx 10^{-3} \left(\frac{R}{T}\right)^4\,.
                   \label{eq:Wave}
\end{equation}
Non-relativistic motion requires $T \gg R$. If, for example, we take
$R/T \approx 1/10$, we would get the estimate  $\widetilde{W}_R
\approx 10^{-7}$.  However, it is plausible that a treatment which
allows for relativistic motion of the electrons would yield a larger
result, perhaps approaching the limiting value of  $\widetilde{W}_R
\approx 10^{-3}$ which arises from Eq.~(\ref{eq:Wave}) when $R \approx
T$. This is a topic for future study.

\subsection{Multiple Modes in Empty Space}

Now let us return to the main result of Sect.~\ref{S:multiple_mode}, Eq.~(\ref{E:temp7}), which
describes the effect of a finite bandwidth of excited modes without a cavity. This expression 
may be written in terms of the function $F$ as
\begin{equation}
\widetilde{W}_R = -\frac{\alpha}{2\pi^2}\, \bar{g}(r)\,  \left(\frac{R}{T}\right)^2\,
\frac{\Delta\omega}{\bar{\omega}}\,  F(\bar{\omega} T)\, d\Omega \,.
\end{equation}
Suppose that $\bar{g}(r) \approx -1/3$ and we integrate over a small but finite solid angle
$\Delta \Omega$. Then we have the estimate
\begin{equation}
\widetilde{W}_R \approx  10^{-4}\,  \left(\frac{R}{T}\right)^2\,
 \frac{\Delta\omega}{\bar{\omega}}\,  F(\bar{\omega} T)\, \Delta \Omega \,.
\end{equation}
If we further assume that $\bar{\omega} T \approx 3$, so that $F$ attains its maximum value
of about $10^2$, then we get the estimate
\begin{equation}
\widetilde{W}_R \approx  10^{-2}\,  \left(\frac{R}{T}\right)^2\,
 \frac{\Delta\omega}{\bar{\omega}}\,   \Delta\Omega \,.
\end{equation}
All of the factors in the above expression, $R/T$, ${\Delta\omega}/{\bar{\omega}}$, and
$\Delta \Omega$, should be small compared to unity for our analysis to
be strictly valid. If we take all three of these factors to be of
order $10^{-1}$, then we would obtain  $\widetilde{W}_R \approx
10^{-6}$. Again, it may be possible to do better with an analysis
which removes the restictions on these factors.

\section{Discussion and Conclusions}
\label{sec:final}

Coherent electrons can undergo decoherence due to coupling to the quantized electromagnetic field,
even if no real photons are initially present. The effect can be given two complementary 
descriptions in terms of either a fluctuating Aharonov-Bohm phase, or of photon emission.
In general, the presence of real photons increases the degree of decoherence. However, as we
have seen, it is possible to temporarily decrease the decoherence if the photons are in a
squeezed vacuum state. This recoherence requires that the electrons be selected to pass through
the interferometer in the correct phase relative to the excited mode or modes of the 
electromagnetic field. An interference pattern formed from such selected electrons can have
a slightly increased contrast compared to the case where no photons are initially present. 
This be interpreted as a transient suppression of Aharonov-Bohm phase fluctuations, analogous
to the suppression of vacuum fluctuations which can lead to negative energy densities. Just
as there are quantum inequalities which limit negative energy density, we have found limits
on the amount of recoherence possible in a squeezed vacuum state. Although the recoherence effect
is small, it may be large enough to be observable. 

\begin{acknowledgments}
We would like to thank C.I. Kuo, D.S. Lee, K.W. Ng and C.H. Wu for useful discussions. This 
work was supported in part by the National Science Foundation under Grant PHY-0555754.
\end{acknowledgments}

\appendix

\section{properties of the squeezed vacuum state}
\label{App:squeezed_state}
The single-mode squeezed vacuum state $\left|\zeta\right>$ is defined by
\begin{equation*}
    \left|\zeta\right>=S\left(\zeta\right)\left|0\right>\,,
\end{equation*}
where the squeeze operator $S\left(\zeta\right)$ is~\cite{Stoler,Caves}
\begin{equation*}
    S\left(\zeta\right)=\exp\left[\frac{1}{2}\left(\zeta^*a^2-
\zeta a^{\dagger2}\right)\right]\,.
\end{equation*}
The operators $a^{\dagger}$ and $a$ are creation and annihilation operators respectively
 satisfying the commutation relation $[a,a^{\dagger}]=1$. The vacuum state 
$\vert0\rangle$ is annihilated by the action of $a$, that is, $a\vert0\rangle=0$. 
The  squeeze parameter $\zeta=r e^{i\theta}$ is an arbitrary complex number with $r$, 
$\theta\in\mathfrak{R}$.

With the help of the operator expansion theorem,
\begin{equation}
    e^{\lambda A}B\,e^{-\lambda A}=B+\lambda\left[A,B\right]+
\frac{\lambda^2}{2!}\left[A,\left[A,B\right]\right]+\cdots\,,
\end{equation}
we readily find for the unitary transformation of the operator $a$
by $S\left(\zeta\right)$,
\begin{equation}
    S^{\dagger}\left(\zeta\right)a\,S\left(\zeta\right)=\mu a-\nu a^{\dagger}\,,
\label{E:form}\\
    \qquad\text{and}\qquad
    S^{\dagger}\left(\zeta\right)a^{\dagger}S\left(\zeta\right)=\mu a^{\dagger}-\nu^*a\,,
\end{equation}
where
\begin{equation}
    \mu=\cosh r\qquad\qquad\nu=e^{i\theta}\sinh r\,,
\end{equation}
and $\mu^2-\left|\nu\right|^2=1$.

The expectation value of $a$ in the squeezed vacuum is given by
\begin{equation}
    \langle\zeta |a |\zeta\rangle=
\langle 0 |S^{\dagger}(\zeta)a\, S(\zeta) |0 \rangle=0
\end{equation}
from Eq.~\eqref{E:form}, and the expectation value of $a^{\dagger}$ is
\begin{equation}
    \left<\zeta\right|a^{\dagger}\left|\zeta\right>=0.
\end{equation}
Moreover, we have
\begin{align}
    &\left<\zeta\right|a^2\left|\zeta\right>=-\mu\nu\\
    &\left<\zeta\right|a^{\dagger2}\left|\zeta\right>=-\mu\nu^*\\
    &\left<\zeta\right|a^{\dagger}a\left|\zeta\right>=\left|\nu\right|^2.\label{E:number}
\end{align}
From Eq.~\eqref{E:number}, it is apparent that the squeezed vacuum
state is not a vacuum state at all, and it has
$\left|\nu\right|^2$ photons on the average.

Next we evaluate the energy density of the electromagnetic fields in a single-mode 
squeezed vacuum state, as well as the total energy. It is assumed that only one of the 
modes of the electromagnetic fields is excited to the squeezed vacuum state while the
 rest of
the modes remain in the vacuum state. This excited mode is denoted
by the wave vector $\bar{\mathbf{k}}$ and polarization $\bar{\lambda}$.
Thus the squeezed vacuum state is created by
\begin{equation}
    \left|\zeta_{\bar{\lambda}\bar{\mathbf{k}}}\right>=
S\left(\zeta_{\bar{\lambda}\bar{\mathbf{k}}}\right)\left|0_{\bar{\lambda}\bar{\mathbf{k}}}\right>\,,
\end{equation}
where $S\left(\zeta_{\bar{\lambda}\bar{\mathbf{k}}}\right)$ is the squeeze
operator for mode $\left(\bar{\lambda},\bar{\mathbf{k}}\right)$,
\begin{equation}
    S\left(\zeta_{\bar{\lambda}\bar{\mathbf{k}}}\right)=
\exp\left[\frac{1}{2}\left(\zeta_{\bar{\lambda}\bar{\mathbf{k}}}^*a^2-
\zeta_{\bar{\lambda}\bar{\mathbf{k}}}a^{\dagger2}\right)\right]\,.
\end{equation}
Here $\zeta_{\bar{\lambda}\bar{\mathbf{k}}}=r e^{i\theta}$ is an arbitrary
complex number. The energy density $\varrho$ of the electromagnetic fields in a 
single-mode squeezed state is given by the expectation value of the corresponding
 energy density operator $\rho$ in the squeezed vacuum state 
$\left|\zeta_{\bar{\lambda}\bar{\mathbf{k}}}\right>$, that is, 
$\varrho(x)=\langle\zeta_{\bar{\lambda}\bar{\mathbf{k}}}|\rho(x)|
\zeta_{\bar{\lambda}\bar{\mathbf{k}}}\rangle$, where the energy density operator is
\begin{equation}
    \rho(x)=\frac{1}{2}\left[E\left(x\right)^2+B\left(x\right)^2\right]\,.
\end{equation}
Let the vector potential $A\left(x\right)$ take the form of the
plane-wave expansion
\begin{equation}
    A\left(x\right)=
\frac{1}{\sqrt{V}}\sum_{\mathbf{k}}\frac{1}{\sqrt{2\omega}}\sum_{\lambda=1}^2
\boldsymbol{\varepsilon}_{\lambda}\left(\mathbf{k}\right)
\left(a_{\lambda k}^{\vphantom{\dagger}}e^{-ik\cdot x}+a_{\lambda k}^{\dagger}e^{ik\cdot x}\right)\,,
\end{equation}
where $\boldsymbol{\varepsilon}_{\lambda}(\mathbf{k})$ is the unit polarization vectors.
$V$ is the normalization volume and $\omega=\left|\mathbf{k}\right|$. The
commutation relations between the creation and the annihilation
operators are
\begin{align}
    \left[a_{\lambda \mathbf{k}}^{\vphantom{\dagger}},a_{\lambda'\mathbf{k}'}^{\dagger}\right]&=
\delta_{\lambda\lambda'}\delta_{\mathbf{k}\mathbf{k}'}\,,\\
    \left[a_{\lambda \mathbf{k}}^{\vphantom{\dagger}},a_{\lambda'\mathbf{k}'}^{\vphantom{\dagger}}\right]
&=\left[a_{\lambda\mathbf{k}}^{\dagger},a_{\lambda'\mathbf{k}'}^{\dagger}\right]=0\,.
\end{align}
Then the electric field $E\left(x\right)$ and the magnetic field
$B\left(x\right)$ are given, respectively, by
\begin{align}
    E\left(x\right)&=
\frac{i}{\sqrt{V}}\sum_{\mathbf{k}}\sqrt{\frac{\omega}{2}}\sum_{\lambda=1}^2
\boldsymbol{\varepsilon}_{\lambda}\left(\mathbf{k}\right)
\left(a_{\lambda\mathbf{k}}^{\vphantom{\dagger}}
e^{-ik\cdot x}-a_{\lambda\mathbf{k}}^{\dagger}e^{ik\cdot x}\right)\,,\\
    B\left(x\right)&=\frac{i}{\sqrt{V}}\sum_{\mathbf{k}}
\frac{1}{\sqrt{2\omega}}\sum_{\lambda=1}^2\mathbf{k}\times
\boldsymbol{\varepsilon}_{\lambda}\left(\mathbf{k}\right)
\left(a_{\lambda\mathbf{k}}^{\vphantom{\dagger}}e^{-ik\cdot x}-
a_{\lambda\mathbf{k}}^{\dagger}e^{ik\cdot x}\right)\,.
\end{align}
Hence the energy density operator is given by
\begin{multline}
    \rho(x)=-\frac{1}{2V}\sum_{\mathbf{k}\,\mathbf{k}'}
\sqrt{\frac{\omega}{2}}\sqrt{\frac{\omega'}{2}}\sum_{\lambda\,\lambda'=1}^2
\Bigl(\boldsymbol{\varepsilon}_{\lambda}\left(\mathbf{k}\right)\cdot
\boldsymbol{\varepsilon}_{\lambda'}\left(\mathbf{k}'\right)+
\left[\boldsymbol{\kappa}\times\boldsymbol{\varepsilon}_{\lambda}
\left(\mathbf{k}\right)\right]\cdot\left[\boldsymbol{\kappa}'\times
\boldsymbol{\varepsilon}_{\lambda'}\left(\mathbf{k}'\right)\right]\Bigr)\\
 \times\left(a_{\lambda\mathbf{k}}^{\vphantom{\dagger}}e^{-ik\cdot x}-
a_{\lambda\mathbf{k}}^{\dagger}e^{ik\cdot x}\right)
\left(a_{\lambda'\mathbf{k}'}^{\vphantom{\dagger}}e^{-ik'\cdot x}-
a_{\lambda'\mathbf{k}'}^{\dagger}e^{ik'\cdot x}\right)\,,
\end{multline}
where $\boldsymbol{\kappa}$ is a unit vector along the direction of the wave
vector. It is straightforward to evaluate the expectation value of the energy
density operator
\begin{align}
    \langle \zeta_{\bar{\lambda}\bar{\mathbf{k}}} |\rho(x)
  |\zeta_{\bar{\lambda}\bar{\mathbf{k}}}\rangle &=
\frac{1}{2V}\left\{\sum_{(\lambda,\mathbf{k})\neq(\bar{\lambda},\bar{\mathbf{k}})}
\frac{\omega}{2}\Bigl(\boldsymbol{\varepsilon}_{\lambda}\left(\mathbf{k}\right)
\cdot\boldsymbol{\varepsilon}_{\lambda}\left(\mathbf{k}\right)+
\left[\boldsymbol{\kappa}\times\boldsymbol{\varepsilon}_{\lambda}
\left(\mathbf{k}\right)\right]\cdot\left[\boldsymbol{\kappa}\times
\boldsymbol{\varepsilon}_{\lambda}\left(\mathbf{k}\right)\right]\Bigr)\right. \nonumber\\
    &\qquad\qquad\qquad+\frac{\bar{\omega}}{2}
\Bigl(\boldsymbol{\varepsilon}_{\bar{\lambda}}\left(\bar{\mathbf{k}}\right)\cdot
\boldsymbol{\varepsilon}_{\bar{\lambda}}\left(\bar{\mathbf{k}}\right)+
\left[\bar{\boldsymbol{\kappa}}\times
\boldsymbol{\varepsilon}_{\bar{\lambda}}\left(\bar{\mathbf{k}}\right)\right]\cdot
\left[\bar{\boldsymbol{\kappa}}\times\boldsymbol{\varepsilon}_{\bar{\lambda}}
\left(\bar{\mathbf{k}}\right)\right]\Bigr) \nonumber \\
    &\Biggl.\qquad\qquad\qquad\qquad\times\left[\left(2
\left|\nu\right|^2+1\right)+\mu\nu\,e^{-2i\bar{k}\cdot x}+\mu\nu^*\,
e^{2i\bar{k}\cdot x}\right]\Biggr\}\,.
\end{align}
We notice that
\begin{equation}
    \boldsymbol{\varepsilon}_{\lambda}\left(\mathbf{k}\right)\cdot
\boldsymbol{\varepsilon}_{\lambda}\left(\mathbf{k}\right)+
\left[\boldsymbol{\kappa}\times
\boldsymbol{\varepsilon}_{\lambda}\left(\mathbf{k}\right)\right]\cdot
\left[\boldsymbol{\kappa}\times\boldsymbol{\varepsilon}_{\lambda}
\left(\mathbf{k}\right)\right]=2\,.
\end{equation}
After subtracting the vacuum contribution, we have the renormalized energy density 
$\varrho_R$ in a squeezed vacuum given by
\begin{align}\label{E:energy}
    \varrho_R\left(x\right)&=\frac{1}{V}\left[\left|\nu\right|^2+
\frac{1}{2}\left(\mu\nu\,e^{-i2\bar{k}\cdot x}+\mu\nu^*\,e^{+i2\bar{k}\cdot x}\right)\right]
\bar{\omega}\notag\\
    &=\frac{1}{V}\,\eta\left[\mu\cos\left(2\bar{k}\cdot x-\theta\right)+
\eta\right]\bar{\omega}\,,
\end{align}
where $\eta=\sinh r$. Note that this can be negative when the condition
$\cos\left(2\bar{k}\cdot x-\theta\right)<0$ is met. Note that the factor which governs 
the sign of $\varrho_R$ is of the same form as $g(r,t)$ defined in Eq.~(\ref{eq:g}).

Accordingly, the renormalized total energy $E_R$ in the squeezed
vacuum state is given by integrating the renormalized energy
density over all quantization volume. If the quantization volume
is sufficiently large, or the periodic boundary conditions are
used for convenience, then the term proportional to
$\cos\left(\cdots\right)$ will vanish and we have
\begin{equation}\label{E:RTE}
    E_R=\int_Vd^3x\,\varrho_R\left(x\right)=\eta^2\bar{\omega}\,.
\end{equation}
The spatial average of the renormalized energy density is then
given by
\begin{equation}\label{E:RTED}
    \bar{\varrho}_R=\frac{E_R}{V}=\frac{1}{V}\,\eta^2\bar{\omega}\,,
\end{equation}
which is always positive.

\end{document}